\documentclass{PoS}

\title{Carpet results on astrophysical gamma rays above 100 TeV}

\ShortTitle{Carpet results on astrophysical gamma rays above 100 TeV}

\author{D.~Dzhappuev, \speaker{S.~Troitsky} and Y.~Zhezher for
the Carpet--3
collaboration\footnote{for collaboration list and acknowledgements see
page~5}\\
Institute for Nuclear Research of the Russian Academy of
Sciences, 60th October Anniversary Prospect 7A, 117312, Moscow, Russia\\
E-mail: \email{st@ms2.inr.ac.ru}}

\abstract{Carpet is an air-shower array at Baksan, Russia, equipped with a
large-area muon detector, which makes it possible to separate primary
photons from hadrons. We report first results of the search for primary
photons with energies $E>100$~TeV.
The
experiment's ongoing upgrade and future sensitivity
are also discussed.}

\FullConference{36th International Cosmic Ray Conference -ICRC2019-\\
		July 24th - August 1st, 2019\\
		Madison, WI, U.S.A.}

\begin{document}

\section{Introduction}
\label{sec:intro}
Gamma-ray astronomy in the PeV domain has several tasks of recognized
importance (for a review, see e.g.\ Ref.~\cite{Lipari}). They include
localization of astrophysical objects and environments where
highest-energy Galactic cosmic rays are accelerated \cite{PeVatrons},
distinguishing~\cite{Murase-gamma, OK-ST-gamma} between Galactic and
extragalactic origin of high-energy astrophysical neutrinos detected by
IceCube \cite{IC}, and now also by Baikal-GVD \cite{Baikal}, and search
for new physics manifesting itself in anomalies of gamma-ray propagation
\cite{ST-axion-rev, RoncadellyPeV}. The method to detect astrophysical
gamma rays of such high energies is determined by two facts: these events
are rare and the primary photon interacts in the upper atmosphere.
Cosmic-ray detectors recording extensive air showers (EAS) are therefore
best suited for the studies, however, special methods are required to
separate events caused by primary gamma rays from huge background of
hadronic cosmic rays. One of the most efficient approaches is to use
separate detectors for electromagnetic and muon components of the EAS:
gamma-induced showers are muon-poor as compared to cosmic-ray induced
ones. Here we report preliminary results of the search for PeV
astrophysical gamma rays by this method.

\section{Experiment and data}
\label{sec:intro}
\subsection{The Carpet--2 experiment}
\label{sec:carpet}
\textit{Carpet--2} is a detector of extensive air showers situated at the
Baksan Neutrino Observatory in Northern Caucasus. It uses the surface
array of scintillator detector stations to record the electromagnetic EAS
component, while the muon component is recorded by its underground
scintillator detector. For a description of the
experiment, see Refs.~\cite{0902.0252, 1511.09397}. The surface detector
determines the arrival direction and time, the core location and the
shower size $N_{e}$, while the 175~m$^{2}$ underground detector determines
the number $n_{\mu}$ of muons which hit the detector. The threshold of
the muon detector is 1~GeV for a vertical muon. We perform Monte-Carlo
simulations of air showers
with CORSIKA~\cite{CORSIKA} (v.~7.5600), using the hadronic-interaction
models QGSJET-II.04~\cite{QGSJET-II.04} and FLUKA-2011.2c~\cite{FLUKA1}.
Artificial events were produced by additional Monte-Carlo simulations of
the installation, recorded and processed in the same way as the real data.
We used two independent data sets described below for the data analysis.

\subsection{Dataset I: $E_{\gamma}>1$~PeV, 1999--2011}
\label{sec:datasetI}
\begin{figure}[tp]
\includegraphics[width=.6\textwidth]{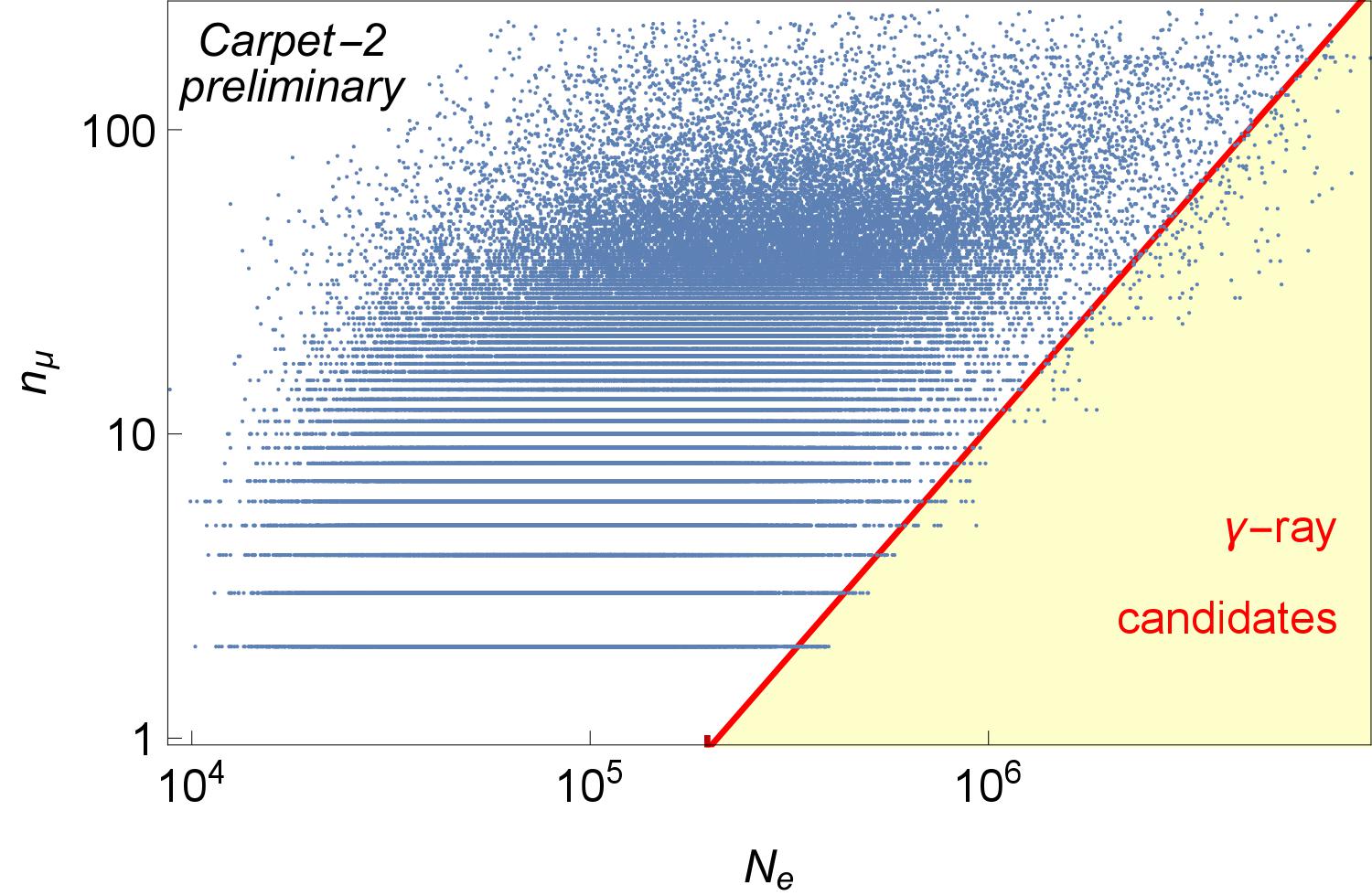}
\caption{Photon candidate region on the $N_{e} - n_{\mu}$
plane, Dataset~I, $E_{\gamma}>1$~PeV (shaded). Blue dots represent data for all events in the
dataset.}
\label{fig:cuts1}
\end{figure}

For the data recorded during the experiment run in 1999--2011 (3080 live
days), the trigger conditions included at least two muons in the
175~m$^{2}$ detector, which is not the optimal cut for the search of
muon-poor photon-induced showers at low energies. Simulations indicate
that, for $E_{\gamma}^{-2}$ primary photon spectrum which we assume
throughout this study, the reconstruction efficiency integrated over the
field of view is $\sim 17\%$ for $E_{\gamma}\sim 1$~PeV. The
efficiency quickly rises with energy and depends on the zenith angle. The
angular resolution of the experiment for primary photons with these
energies is $\sim 4.2^{\circ}$, determined by the Monte-Carlo simulations.
Quality cuts are described in Ref.~\cite{CarpetIC}); 115821 events
passed the cuts.
Additional cuts aimed to separate photon candidate events
on the $N_{e}-n_{\mu}$ plane were determined from simulations to maximize
the efficiency of photon detection and to minimize the hadronic
background; they correspond to the shaded area in Fig.~\ref{fig:cuts1},
\begin{figure}[tp]
\includegraphics[width=.6\textwidth]{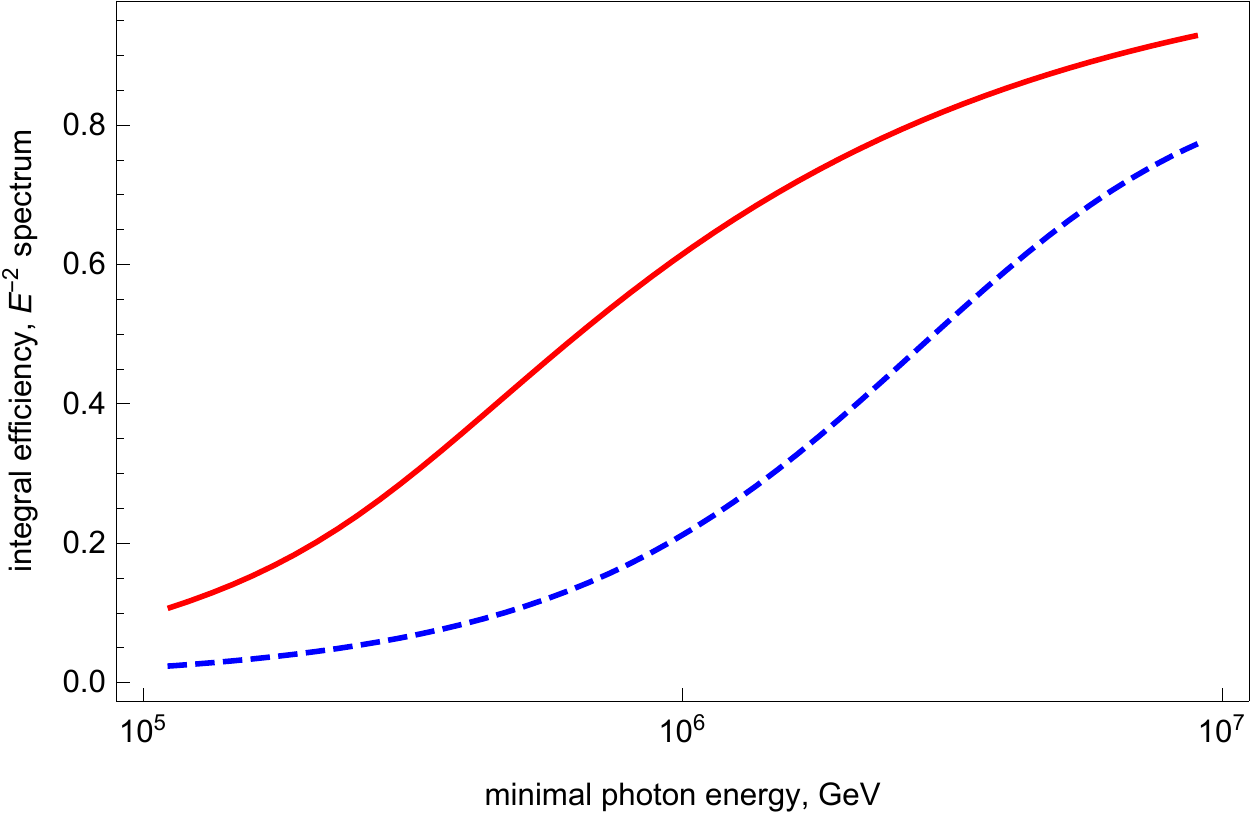}
\caption{Efficiency of detection of gamma rays with $E_{\gamma}>E_{\rm
min}$, assuming $E_{\gamma}^{-2}$ primary spectrum. Blue dashed line:
Dataset~I, red full line: Dataset~II.}
\label{fig:eff}
\end{figure}
\begin{figure}[tp]
\includegraphics[width=.6\textwidth]{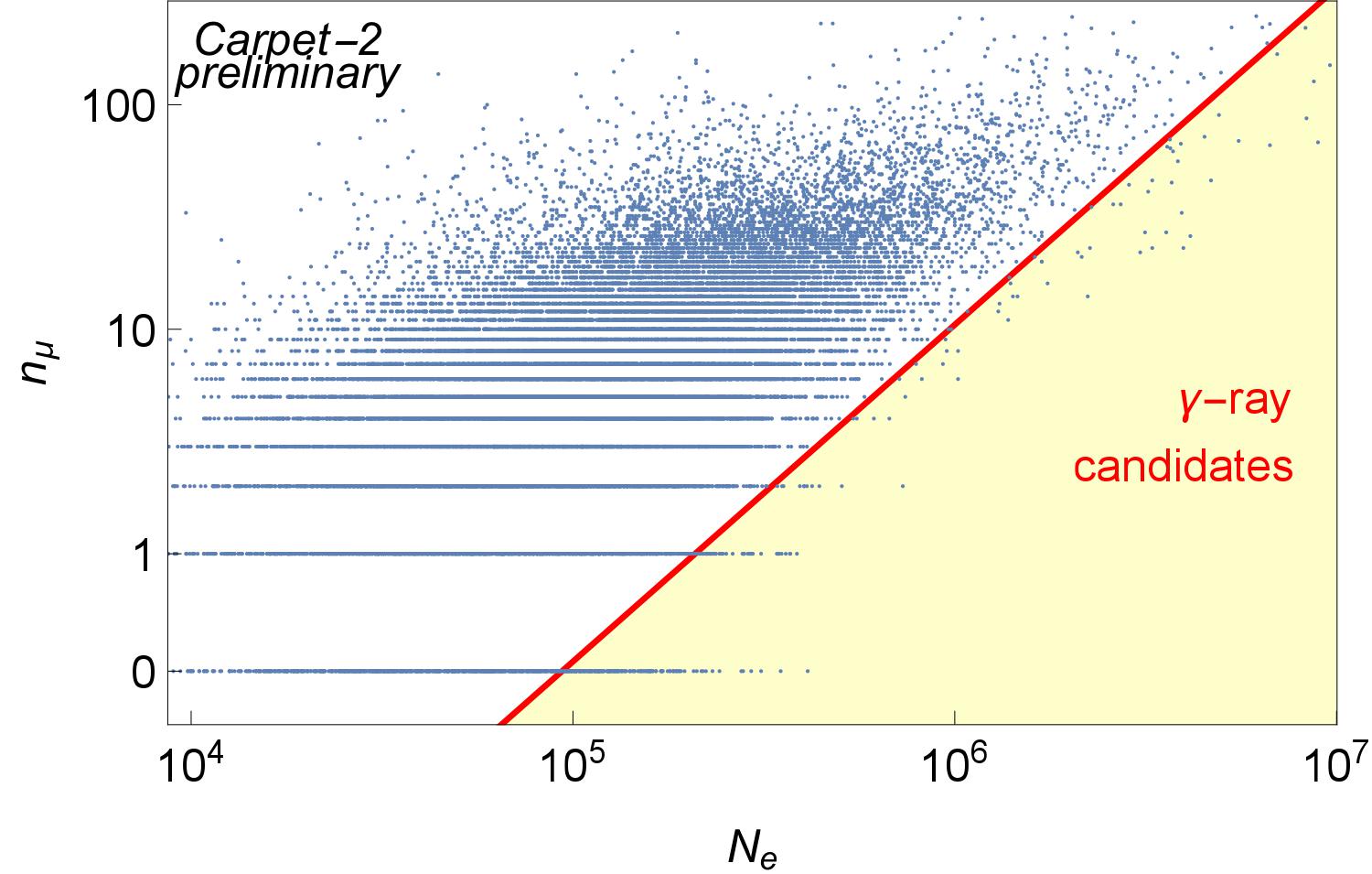}
\caption{Same as Fig.~\ref{fig:cuts1} but for Dataset~II,
$E_{\gamma}>300$~TeV.}
\label{fig:cuts2}
\end{figure}
where the data are also shown. There are 310 photon candidate events in the
data set.

\subsection{Dataset II: $E_{\gamma}>0.3$~PeV, 2018-2019}
\label{sec:datasetII}
Motivated by post-IceCube interest in PeV gamma-ray astronomy, the
experiment was relaunched in 2018 with the new trigger which allowed to
record also events with zero or one muon in the detector. This resulted
in lowering the threshold to $E_{\gamma,{\rm min}}=300$~TeV, with the
efficiency of $\sim 26 \%$. Figure~\ref{fig:eff}
compares the efficiencies
for the two data sets. Here, we report results for 342 live days recorded
in 2018-19; quality cuts were passed for 25876 events. For these trigger
conditions and threshold energy, different photon-candidate cuts are
optimal, see Fig.~\ref{fig:cuts2}.
There are 399 photon candidate events in this data set.

\section{Results}
\label{sec:results}
\subsection{Sky map of excesses}
\label{sec:map}
\begin{figure}[tp]
\includegraphics[width=.6\textwidth]{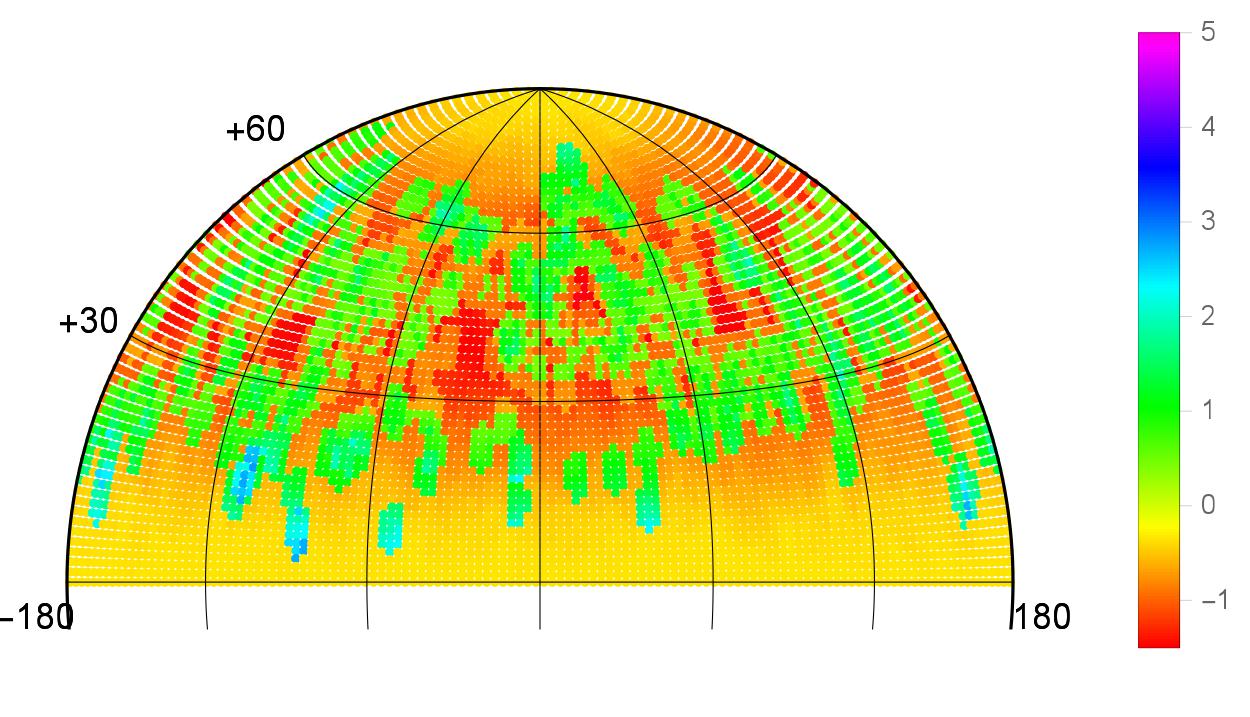}
\caption{Skymap of excesses of counts of events for Dataset~I,
$E_{\gamma}>1$~PeV. The inset presents the colour code of significances,
in standard deviations corresponding to the observed Poisson probability.
Note that the fluctuations are asymmetric because the number of events is
always positive.}
\label{fig:map1}
\end{figure}

\begin{figure}[tp]
\includegraphics[width=.6\textwidth]{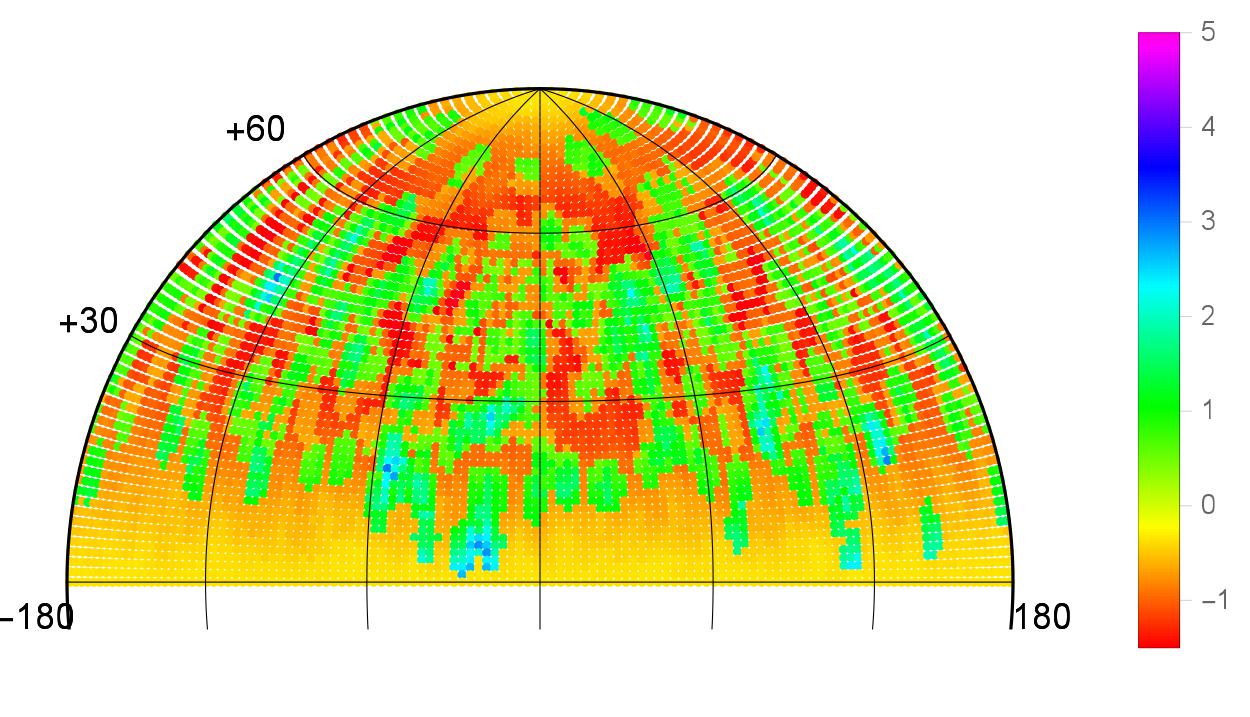}
\caption{Same as Fig.~\ref{fig:map1} but for Dataset~II,
$E_{\gamma}>300$~TeV.}
\label{fig:map2}
\end{figure}

Point sources of gamma rays are expected to reveal themselves on the
isotropic background, which includes both diffuse photons and
possible hadrons contaminating the data set. Assuming isotropy, one
calculates the expected background for any given direction in the sky and
compares with the real number of observed events from this direction,
taking into account the angular resolution. A blind search for excesses
may be performed with excess sky maps presented in Fig.~\ref{fig:map1}
and Fig.~\ref{fig:map2} for two datasets.

\subsection{Predefined sources}
\label{sec:sources}
We have defined in advance the list of four sources to be monitored, in
order to avoid additional, hard to calculate, corrections to the
significance related to the ``look-elsewhere effect''. Preliminary results
of the analysis above 1~PeV have been reported in
Ref.~\cite{VLVNT} (see also Ref.~\cite{CarpetIC} for PeV photons
from the directions of IceCube events); here we refine these results and
present also those for Dataset~II of lower energies, see
Table~\ref{tab:limits}.
\begin{table}
\begin{tabular}{ccccccc}
\hline
\hline
Source & \multicolumn{3}{c}{Dataset I: $E_{\gamma}>1$~PeV} &
\multicolumn{3}{c}{Dataset II: $E_{\gamma}>0.3$~PeV}  \\
name   &
expected & observed & flux, cm$^{-2}$\,s$^{-1}$ &
expected & observed & flux, cm$^{-2}$\,s$^{-1}$ \\
\hline
Crab    & 0.69  & 0 & $<4.69\times 10^{-13}$
        & 0.95  & 4 & $<1.57\times 10^{-11}$\\
Cyg~X-3 & 1.75  & 2 & $<3.00\times 10^{-13}$
        & 2.08  & 1 & $<1.83\times 10^{-12}$\\
Mrk~421 & 1.59  & 5 & $<6.32\times 10^{-13}$
        & 1.90  & 4 & $<5.45\times 10^{-12}$\\
Mrk~501 & 1.70  & 0 & $<8.8\times 10^{-14}$
        & 1.93  & 1 & $<2.00\times 10^{-12}$\\
\hline
\hline
\end{tabular}
\caption{Results of the photon search from four predefined sources. Flux
upper limits are 95\% CL.}
\label{tab:limits}
\end{table}
Interestingly, Mrk~421 demonstrates a weak, $\sim (2-3) \sigma$, excess in
both independent data sets (for Dataset~I, this was pointed out in
Ref.~\cite{VLVNT}). Gamma rays of the energies considered here should not
arrive from extragalactic objects because of intense pair production on
the cosmic microwave background~\cite{Nikishov}. With rather poor angular
resolution of {\it Carpet-2}, one cannot exclude that these photons arrive
from a different, Galactic, source, though a candidate source is unknown.
Alternatively, this may be a statistical fluctuation. A more detailed
study of this direction in the sky will be reported elsewhere.

\section{Conclusions and outlook}
\label{sec:concl}
This work reports preliminary results on the search of very-high-energy
cosmic gamma rays with \textit{Carpet-2}, a EAS detector at the Baksan
Neutrino Observatory. Thanks to a 175~m$^{2}$ muon detector, separation of
photon-induced showers from the bulk of cosmic-ray induced events becomes
possible. We consider two datasets, Dataset~I with large exposure and
trigger conditions allowing for the search of $E_{\gamma}>1$~PeV photons
and Dataset~II with smaller exposure but lower gamma-ray energy threshold,
$E_{\gamma}>0.3$~PeV. We present upper limits on the photon fluxes from a
set of four predefined point sources. One of the sources, Mrk~421,
demonstrates a weak excess in both independent data sets;  its origin
remains to be understood. Crab nebula demonstrates an insignificant excess
at lower energies, Dataset~II, and will be monitored. More results will be
discussed at the conference.

The installation is being upgraded to
\textit{Carpet--3}, with the extended
410~m$^{2}$ muon detector. The new muon detector is already
installed and it is planned to start data taking in the new configuration
this fall, which should result in a crucial improvement in the gamma-hadron
separation. Additional surface-detector stations will be installed to
increase the collecting area. Preliminary simulations~\cite{VLVNT} allow
to hope that one year of \textit{Carpet--3} live data taking might probe
Galactic models of IceCube neutrinos~\cite{Semikoz} with the diffuse
photon flux at $E_{\gamma}>100$~TeV.

\section*{Collaboration list}
\sloppy
D.\,D.\,Dzhappuev\footnote{Institute for Nuclear Research of the Russian
Academy of
Sciences, 60th October Anniversary Prospect 7A, 117312, Moscow, Russia},
I.\,M.\,Dzaparova$^{1}$\footnote{Institute of Astronomy,
Russian Academy of Sciences, 119017, Moscow,
Russia},
E.\,A.\,Gorbacheva$^{1}$,
I.\,S.\,Karpikov$^{1}$,
M.\,M.\,Khadzhiev$^{1}$,
N.\,F.\,Klimenko$^{1}$,
A.\,U.\,Kudzhaev$^{1}$,
A.\,N.\,Kurenya$^{1}$,
A.\,S.\,Lidvansky$^{1}$,
O.\,I.\,Mikhailova$^{1}$,
V.\,B.\,Petkov$^{1,2}$,
V.\,S.\,Romanenko$^{1}$,
G.\,I.\,Rubtsov$^{1}$,
S.\,V.\,Troitsky$^{1}$,
A.\,F.\,Yanin$^{1}$,
Ya.\,V.\,Zhezher$^{1}$,
K.\,V.\,Zhuravleva$^{1}$.

\section*{Acknowledgements}
Experimental work of \textit{Carpet--2} is performed in the
laboratory of ``Unique Scientific Installation -- Baksan Underground
Scintillating Telescope''at the ``Collective Usage Center -- Baksan
Neutrino Observatory of INR RAS'' under support of the Program of
fundamental scientific research of the RAS Presidium ``Physics of
fundamental interactions and nuclear technologies''.
The work of a part of the group
(DD, EG, MKh, NK, AUK, ANK, AL, OM, AY) on the upgrade of the
installation was supported in part by the RFBR grant 16-29-13049. The work
of DD, ID, AUK, ANK, AL and VP on PeV gamma rays from Crab
was supported in part by the RFBR grant
16-02-00687.
The work of ST and KZ on PeV gamma rays from extragalactic sources
(Mrk~421, Mrk~501) in the context of constraining the anomalous
transparency of the Universe was supported by the Russian Science
Foundation, grant 18-12-00258. Monte-Carlo simulations have been
performed at the computer cluster of INR Theoretical Physics Department.

\end{document}